\documentclass[twocolumn,showpacs,preprintnumbers,amsmath,amssymb]{revtex4}

\usepackage{graphicx}
\usepackage{epsfig}
\usepackage{dcolumn}
\usepackage{bm}

\begin{document}

\title {Scaling in the Lattice Gas Model}

\author{F.Gulminelli$^{(1)}$, Ph.Chomaz$^{(2)}$, M.Bruno$^{(3)}$, 
                      M.D'Agostino$^{(3)}$ }

\affiliation{
(1) LPC (IN2P3-CNRS/ISMRA et Universit\'{e}), F-14050 Caen c\'{e}dex, France \\
(2) GANIL (DSM-CEA/IN2P3-CNRS), B.P.5027, F-14021 Caen c\'{e}dex, France \\
(3) Dipartimento di Fisica and INFN, Bologna, Italy }

\begin{abstract}
A good quality scaling of the cluster size distributions is obtained for the
Lattice Gas Model using the Fisher's ansatz for the scaling function. This
scaling identifies a pseudo-critical line in the phase diagram of the model that
spans the whole (subcritical to supercritical) density range. The independent
cluster hypothesis of the Fisher approach is shown to describe correctly the
thermodynamics of the lattice only far away from the critical point.
\end{abstract}


\maketitle

Since the first heavy ion experiments multifragmentation has been tentatively
connected to a critical phenomenon \cite{historique}. The recent determination
of a consistent set of critical exponents in different multifragmentation data
\cite{eos,michela} tends to confirm this hypothesis even if the finite size
corrections to scaling are largely unknown. On the other side the experimental
observation of a flattening of the caloric curve \cite{caloric,natowitz} and the
measurement of a negative heat capacity \cite{palluto} points towards a
first order phase transition as it is also suggested by the thermodynamics of
statistical multifragmentation models \cite{gross}. 
The debate on the order of the transition has been further animated by 
a very recent analysis of the Isis data \cite{moretto2} which 
shows a high quality scaling of the fragment size distribution over 
a wide range of charges and deposited energies with an ansatz for the 
scaling function taken from the Fisher droplet model \cite{fisher} which approximates a real fluid as an ideal gas of clusters. 
The critical temperature extracted from the Fisher analysis is identified as the
temperature of the thermodynamical critical point and the whole coexistence line
of the nuclear phase diagram is reconstructed under the hypothesis that the
Fisher model gives a good description of the multifragmentation phenomenon
\cite{moretto2}.

In this paper we apply the Fisher scaling method of ref.\cite{moretto2} to the
Lattice Gas model, which is a well known paradigm of first as well as second
order phase transitions \cite{lee}, the canonical version of this model being
isomorphous to the Ising model at fixed magnetization. 
We will show that the observation of Fisher
scaling does not allow to determine the location of the critical point. Critical
behaviors are observed on a line at supercritical density \cite{campi} which
extends inside the coexistence region \cite{noi,richert} 
due to finite size effects.
We suggest that the inadequacy of the Fisher model 
to describe the lattice gas phase diagram comes from the Fisher 
independent cluster hypothesis which should be correct only at low
densities and high temperatures ; indeed we demonstrate that the thermodynamics reconstructed from the Fisher partition sum coincides with the exact one only in the low density and high temperature regime.  

In our implementation of 
the lattice gas model \cite{lee} the $N$ sites  of a cubic
lattice are characterized by an occupation number $n_i$ which is
defined as $n_i =0 (1) $ for a vacancy (nucleon).
Particles occupying nearest neighboring sites interact with a 
constant coupling $c$. 
The relative particle density $\rho/\rho_0$ 
is defined as the number of occupied sites
divided by the total number of sites and is linked to the  magnetization of the Ising model by $\rho/\rho_0=2m-1$. 
Observables expectation values are evaluated in the canonical ensemble sampled
through a standard Metropolis algorithm \cite{lattice_noi}. 
The use of the canonical constraint allows 
a direct exploration of the coexistence region \cite{richert}. 
The backbending of the chemical potential as a function of the particle number and of the pressure as a function of the density allows an unambiguous definition of the phase diagram even for a finite system \cite{noi}. 
Fragments are defined within the standard Coniglio-Klein bond breaking
probability between occupied neighbor sites. For all technical details, see
refs.\cite{noi,lattice_noi}.

\begin{figure}

\includegraphics*[height=0.9\linewidth]{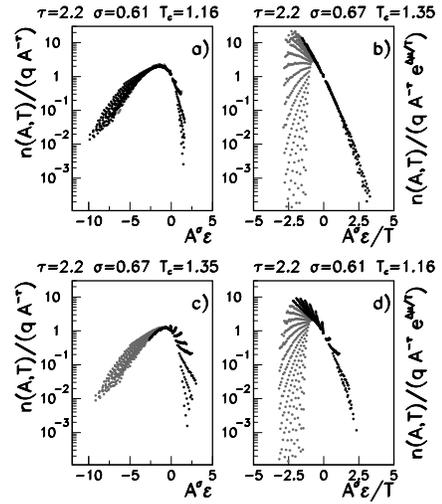}

\caption{RG scaling from eq.\ref{scaling}(figs.1a,1c) 
and Fisher scaling from eq.\ref{fisher}
(figs.1b,1d) of the cluster size distribution in a 8X8X8 cubic lattice at the
critical density for temperatures $0.36<T/c<3.6$ 
and cluster sizes $4<A<30$. The critical
parameters corresponding to each figure are indicated.}

\end{figure}

Renormalization group arguments lead 
to the expectation that in the vicinity of
the critical point $(T_c,\rho_c)$ 
the size distribution scales as \cite{stauffer}
\begin{equation}
  n(A,T) = q A^{-\tau }f\left( \epsilon A^{\sigma }\right)
\label{scaling}
\end{equation}
where $\epsilon$ measures the distance from the critical point 
$\epsilon=(T_c-T)/T_c$, $q$ is a normalization constant, 
$f$ is the scaling function and $\tau,\sigma $ are critical exponents.
We will refer to eq.(\ref{scaling}) in the following as to RG scaling.  
The RG scaling analysis performed on a $8X8X8$ lattice with periodic boundary
conditions at the critical density $\rho_c=1/2$ is 
displayed in figure 1a\cite{noi,lattice_noi}. A good scaling behavior is
observed for all temperatures $0.36<T/c<3.6$ and all cluster sizes $4<A<30$.
The critical exponents $\tau=2.2, \sigma=0.61$ are close to the expected values of the liquid-gas universality class $\tau=2.2, \sigma=0.64$ and the critical temperature $T_c = 1.16c$ is in good agreement both with the
temperature at the thermodynamical critical point \cite{noi} $T_c^{th}=1.22c$
and with the expected critical temperature at the thermodynamical limit $T_c^\infty = 1.12c$. Indeed finite size
corrections to scaling have been evaluated \cite{lattice_noi} and found to be
small. The method used to extract the critical parameters is discussed in detail
in refs. \cite{noi,lattice_noi}.

In the Fisher droplet model \cite{fisher} the vapor coexisting with a liquid in
the mixed phase of a liquid-gas phase transition is schematized as an ideal
gas of clusters. A similar scaling around
the critical point 
is supposed by this model  but a different form is suggested for the
scaling function
\begin{equation}
 n(A,T) = q A^{-\tau } exp(\frac{A \Delta\mu-c_0\epsilon A^\sigma}{T})
\label{fisher}
\end{equation}
Here $\Delta\mu$ represents the 
difference in chemical potential between the two
phases, and  $c_0$ is the surface energy coefficient. Since both $\Delta\mu$ 
and $c_0$ can be in principle temperature dependent, we have parametrized these
quantities as polynomials of order 4 and 1 respectively following 
ref.\cite{moretto1}; the normalization $q$ has been taken as in the
infinite system \cite{elliott}.  
The critical parameters obtained from the best $\chi^2$ fit as well as the 
scaled distributions are shown in figure 1b. Only temperatures lower than the
maximum production temperature for each size have been used in the fit (black
dots in figure 1). The scaling is violated only for higher temperatures (grey
dots) consistently with the Fisher approach which modelizes only the vapor
coexisting with a liquid, i.e. is relevant for temperatures $T<T_c$. 
It is surprising that 
two such different ansatz for the scaling function lead to a comparable quality
for the scaling of the size distributions and to coherent and 
close values for the critical exponents; this remarkable result confirms the
wide universal validity of generic thermal scalings \cite{ther_sca}.
The main ambiguity concerns the critical temperature
which comes out about 20\% higher with the Fisher technique. 
This difference is not a
broadening effect due to finite sizes, a well defined critical point being
replaced in small systems by a wide spread critical region. As a
matter of fact, if the RG critical parameters of figure 1a are implemented in
the Fisher analysis and a reduced 7 parameters fit is done with the ansatz
(\ref{fisher}) the scaling is clearly violated even at low temperatures (figure
1d) and the same thing is true if the Fisher scaling parameters of figure 1b are
inserted in the RG ansatz eq.(\ref{scaling}) as shown by figure 1c. This
means that the two scaling ansatz are not equivalent and the good quality
of the scaling is insufficient to prove the adaptation of the model to the
data. Therefore before giving a physical meaning to the precise value of the 
extracted critical temperature one should a priori know
if the chosen scaling ansatz is consistent with the system under study. 
On the other side the critical exponents seem to be very robust and depend
only very slightly on the scaling hypothesis.

\begin{figure}

\includegraphics*[height=0.9\linewidth]{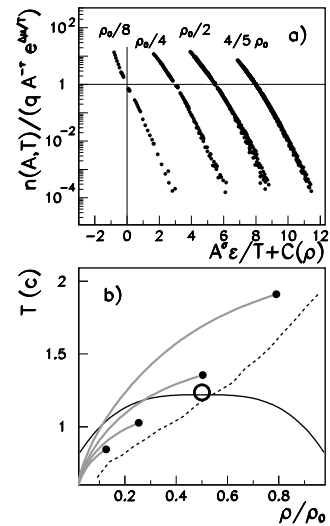}

\caption{Fig.2a: Fisher scaling as in fig.1b but at different densities.
Fig.2b: thermodynamical coexistence line (full line) 
and region of critical partitions (dashed
lines) from ref.[\cite{noi}]. 
Grey lines: coexistence line reconstructed from fragment
partitions via eq.(3).}

\end{figure}

The analysis shown in figure 1 was performed at a constant density equal to the
critical density.In the case of nuclear collisions it is not obvious that
multifragmentation occurs systematically at the critical density. In particular
the good adequacy between statistical models and data favours a lower value for 
the freeze out density \cite{gross} and similar information come  from
interferometry measurements\cite{aladin}. On the other side recent calculations
in the framework of classical molecular dynamics \cite{campi} propose an early
fragmentation at supercritical density. To understand the effect of volume we
have performed different lattice gas calculations at different densities.
As shown in the upper part of figure 2, a very good scaling is observed for all
subcritical as well as supercritical densities. In all cases the values of the 
critical exponents are comparable but the critical temperature is a monotonically increasing function of the density. 
To visualize all the results on the same picture 
a constant horizontal shift $C(\rho)$ is given to each scaled distribution. 
The critical temperatures obtained for each density 
are represented by the black symbols
in the lower part of figure 2. In this figure the full line gives the
coexistence line of the model calculated in a precise way from the derivatives
of the canonical partition sum \cite{noi}. The locus of criticality lies
approximately over a line which passes close to the thermodynamical critical
point (open dot) but extends further at supercritical 
(Kertesz line\cite{campi}) as well as subcritical
densities inside the coexistence region. A qualitatively similar behavior has
been already observed with the RG analysis (dashed line in figure 2b)
\cite{noi} and has been interpreted in terms of finite size effects.
This result implies that in the framework of the lattice gas model the
observation of Fisher scaling and more generally of a critical behavior does not
allow to localize the critical point and is compatible also with
fragment formation at low density inside the coexistence region.

A first order phase transition in a finite system corresponds to a concavity
anomaly in the free energy $F=TlnZ$ which in turn leads to a backbending of
the canonical chemical potential 
$\mu = f + \rho \partial_\rho f$ where $f$ is the free energy per particle.
The coexistence line in figure 2b corresponds to the equality of the
chemical potentials $\Delta \mu = 0$ on the liquid and gas branch defined by a
Maxwell construction \cite{noi,topology}. If the Fisher model is a good
approximation to the Lattice Gas physics it should be 
therefore possible to reconstruct
the vapor side of the coexistence line $\rho_{CL}(T)$ 
directly from the fragment yields as proposed in
ref.\cite{moretto2} as
\begin{equation}
\rho_{CL} = \sum_A n(A,T) A exp(-A \Delta \mu)
\label{coex_line}
\end{equation}
where the sum extends over all fragments but the biggest. The resulting curves
are given by the grey lines in figure 2b for the four different densities
shown in figure 2a. The end point of the lines giving by
construction the total density of the system 
and the critical temperature extracted by
the Fisher fit, these lines are obviously meaningless if the system is
fragmenting at a density different from the critical density $\rho_c=1/2$. 
However even at $\rho=\rho_c$ when the
thermodynamical critical point is included in the data set 
the reconstruction of the coexistence line is very poor.
In particular the curvature of this line at the reconstructed critical
temperature corresponds to an exponent $\beta=0.84$ which 
strongly deviates from the expected $\beta=0.31$ 
exponent given by the thermodynamical coexistence line (full line in fig.2b) 
which would be consistent with the liquid gas universality 
class ($\beta=0.33$)and with the critical exponents extracted from the clusters. 

\begin{figure}

\includegraphics*[height=0.9\linewidth]{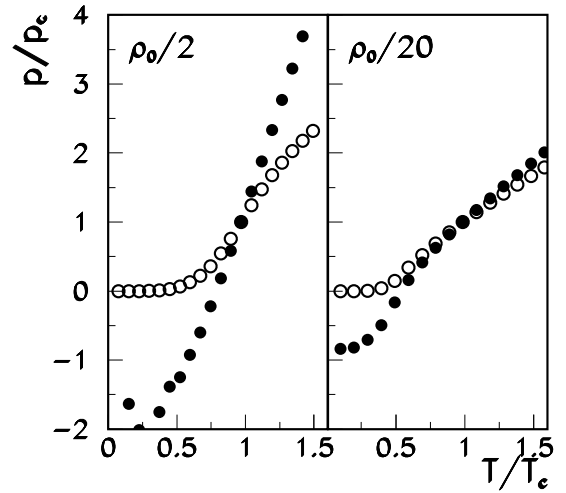}

\caption{normalized pressure versus temperature at two different densities
from the exact canonical partition sum (full dots) and from the ideal gas
approximation eq.(4) (open dots).}

\end{figure}

This means that despite the magnificent scaling shown by figure 2a 
the physics of the Fisher droplet model does not correspond to the Lattice gas.

This may look surprising since the main hypotheses of the Fisher model are
shared by the Lattice : clusters are essentially defined
by a volume and a surface contribution; they exhibit a critical behavior at the
thermodynamical critical point; the statistical weight of a given configuration
is given by a Boltzmann factor.
The spectacular collapse of all the cluster distributions on the single curve 
of figure 2 indeed indicates that the Fisher ansatz gives a good prediction 
of an important part of the physics of the Lattice, i.e. the inclusive yields. 
However it may be interesting to remark that a somewhat different value for 
$\beta$ can be obtained by changing the order of the polynomial assumed for 
$\Delta \mu$ without any sizeable change in the quality of the scaling. This 
suggests that the informations contained in the inclusive yields may be 
insufficient to pin down the thermodynamics of the system.  

The two models strongly differ on one point:
if in the Fisher picture fragments constitute an ideal vapor of non interacting
composite particles (the individual production probabilities are factorized) in
all microscopic models as the Lattice Gas, interactions among fragments are
naturally taken into account through the volume they exclude and through the
surface coupling between neighboring fragments.
One may therefore wonder if these interactions, which seem to affect in a 
non crucial way the inclusive yields, 
may induce important differences in the thermodynamics.
The importance of these effects can be studied by testing the deviation of the
Lattice equation of state from the ideal gas hypothesis of the Fisher model.
If fragments can be modelized as an ideal classical gas, in a
constant volume transformation the pressure can be deduced directly from the
fragments yields \cite{moretto2} 
\begin{equation}
\frac{p}{p_c}=\frac{\sum_A n(A,T)}{\sum_A n(A,T_c)}\frac{T}{T_c}
\label{ideal_gas}
\end{equation}
where the sum extends over the vapor phase (all fragments but the biggest) and
$T_c=T_c(\rho)$ is the temperature obtained from the Fisher fit for each density
(black dots in figure 2b). This pressure is shown by the open dots in figure 3
at two different densities and can be compared to the exact pressure of the
model from ref.\cite{lattice_noi} $p=\rho^2 \partial_\rho f$ (black dots).
Not surprisingly, the gas of clusters behaves as an ideal gas only at low
density and high temperature. The attractive interaction among fragment surfaces
causes the pressure to become negative at low temperatures (the system is bound)
while an extra pressure comes from the excluded volume interaction at high
density.

In conclusion in this paper we have analyzed the fragment size 
distributions issued
of the canonical implementation of the Lattice Gas model 
by means of Fisher scaling.
A very good scaling is observed at subcritical as well as supercritical
densities with values for the critical
exponents compatible (within finite size effects) with the universality class of
the model. This implies that the observation of scaling does not allow to infer
the position of the critical point and is compatible with a fragmentation 
inside the coexistence region of a first order phase transition. 
Knowing that the scaling function of the model is sensibly different from the
Fisher exponential ansatz, 
Fisher scaling appears as a very powerful tool to extract
critical exponents in a way which is essentially independent of the assumptions
made on the detailed shape of the scaling function.
On the other side the reliability of thermodynamical 
quantities extracted from the Fisher analysis
(coexistence curve, critical temperature, saturated pressure..) 
for any set of experimental or simulated data depends critically on the
possibility of approximating the fragment partitions as an ideal noninteracting
gas. In the case of the Lattice Gas model this hypothesis is verified
only for densities much lower and temperatures much higher than the ones of the 
critical point.

This work has been partially supported by NATO grants CLG 976861.


\end{document}